\documentclass[manuscript]{aastex}

\shorttitle{State transitions in a ULX?}
\shortauthors{Godet et al.}

\begin{document}


\title{First evidence for spectral state transitions in the ESO243-49 hyper
luminous X-ray source HLX-1}


\author{O. Godet, D. Barret, N. A. Webb}
\affil{Universit\'e de Toulouse, UPS, CESR, 9 Avenue du Colonel Roche, F-31028 Toulouse Cedex 9, France}

\and 

\author{S. A.  Farrell}
\affil{
Department of Physics and Astronomy, University of Leicester, University
Road, Leicester, LE1 7RH, UK}

\and 

\author{N. Gehrels}
\affil{NASA/Goddard Space Flight Center, Greenbelt, MD 20771, USA}

\begin{abstract}

The brightest Ultra-Luminous X-ray source (ULX), ESO 243-49 HLX-1, with a 0.2
- 10 keV X-ray luminosity of up to 10$^{42}$ erg s$^{-1}$, provides the
strongest evidence to date for the existence of intermediate mass black
holes. Although small scale X-ray spectral variability has already been
demonstrated, we have initiated a monitoring campaign with the X-ray Telescope
onboard the {\it Swift} satellite to search for luminosity-related spectral
changes and to compare its behaviour with the better studied stellar mass
black holes.  In this paper, we report a drop in the XRT count rate by a
factor of $\sim 8$ which occurred simultaneously with a hardening of the X-ray
spectrum. A second observation found that the source had re-brightened by a
factor of $\sim 21$ which occurred simultaneously with a softening of the
X-ray spectrum. This may be the first evidence for a transition between the
low/hard and high/soft states.

\end{abstract}


\section{Introduction}

The most compelling evidence for the existence of intermediate mass black
holes (IMBHs) comes from the observation of ULXs, extragalactic X-ray sources
with bolometric luminosities exceeding 10$^{39}$ erg s$^{-1}$ which are
located outside the nucleus of the host galaxy. These X-ray luminosities -- if
they are assumed to be isotropically radiated -- are up to several orders of
magnitude above the Eddington limit for the maximum mass of stellar mass black
holes \citep[e.g.][]{Roberts07}. There is still an open debate about whether
any ULXs host IMBHs. How IMBHs form and evolve is also a subject of intense
debate, but they are thought to be associated with events such as the
implosion of massive stars formed during the very first stages of star
formation, the collapse of dense star clusters, and the early growth of
supermassive black holes lying in the center of galaxies \citep{miller09}.

2XMM J011028.1--460421, referred to hereafter as HLX-1, was discovered
serendipitously by XMM-Newton on 23 November 2004 (hereafter XMM1) in the
outskirts of the edge-on spiral galaxy ESO 243--49, at a redshift of 0.0224
\citep{Afonso05}. Its 0.2 -- 10 keV unabsorbed X-ray luminosity, assuming
isotropic emission, was found to be $1.1 \times 10^{42}$ ergs s$^{-1}$: one
order of magnitude larger than the previously known brightest ULX (Miniutti et
al. 2006). A second XMM-Newton observation performed 4 years later (on 28
November 2008 -- hereafter XMM2) revealed that the source spectrum and X-ray
luminosity had changed. From its highest X-ray luminosity and the conservative
assumption that the observed value exceeded the Eddington limit by a factor of
10, a lower limit of 500 M$_\odot$~was derived for the black hole (BH) in
HLX-1, thus providing the strongest evidence so far that IMBHs do exist
\citep{Farrell09}.

The evidence for spectral variability in HLX-1 from the two XMM-Newton
observations was the motivation for monitoring the source with the {\it Swift}
X-ray Telescope \citep[XRT;][]{Burrows05}. The aim of this program is to study
its spectral properties as a function of its X-ray luminosity to search for
any of the three canonical X-ray states identified in Galactic black hole
binaries (GBHBs) \citep[see for instance][]{MR06}. Observing transitions
between these states could thus provide important information on the physical
nature of ULXs \cite[e.g.][]{Kajava:2009sj}.  Despite its lower effective area
compared to XMM-Newton and Chandra, the XRT can record enough counts in
relatively short exposure times (few to tens of kiloseconds) to track any
significant luminosity and spectral changes. This has been recently
demonstrated by Kaaret \& Feng (2009) who reported on a monitoring campaign
with the XRT of three prototype ULXs, i.e. Holmberg IX X-1, NGC 5408 X-1 and
NGC 4395 X-2. No spectral state changes were observed in any of these three
sources, despite significant changes in their X-ray luminosities (by up to a
factor of 9).

In this paper, we present the results of the first two observations of our
on-going monitoring campaign of HLX-1 with the {\it Swift} XRT. Compared with
a previous XRT pointing performed one year before, this observation revealed
not only a dramatic drop in the XRT count rate (by a factor of $\sim 8$) but
also a significant hardening of the X-ray spectrum as well as a dramatic
re-brightening (by a factor $\sim 21$) occurring simultaneously with a clear
softening of the X-ray spectrum.

\section{X-ray observations}

HLX-1 was observed with {\it Swift} under the Target-of-Opportunity (ToO)
program on 3 occasions so far: S1 $=$ 2008-10-24 (33.5 ks), S2 $=$ 2009-08-05
(19.2 ks) and S3 $=$ 2009-08-16 (19.1 ks). All the {\it Swift}-XRT Photon
Counting data were processed using the tool {\scriptsize XRTPIPELINE}
v0.12.3\,\footnote{See {\scriptsize
http://heasarc.gsfc.nasa.gov/docs/swift/analysis/}}. We used the grade 0-12
events, giving slightly higher effective area at higher energies than the
grade 0 events, and a 20 pixel (47.2 arcseconds) radius circle to extract the
source and background spectra using {\scriptsize XSELECT} v2.4a. The
background extraction region chosen close to the source extraction region is
the same for the three epochs S1, S2 and S3. No XMM-Newton sources are present
inside the background extraction region. The average background count rate is
$6.2\pm 2.2\times 10^{-4}$ count s$^{-1}$.  The ancillary response files were
created using {\scriptsize XRTMKARF} v0.5.6 and exposure maps generated by
{\scriptsize XRTEXPOMAP} v0.2.5. We fitted all the spectra within {\scriptsize
XSPEC} v12.5.0a using the response file {\scriptsize
SWXPC0TO12S6$_{-}$20070901V011.RMF}, which includes an improvement of the soft
energy response of the XRT \citep{Godet09a, Godet09b}; which is essential for
a source as soft as HLX-1.

The XRT monitoring revealed that HLX-1 was highly variable over the past 10
months with a drop in count rate by a factor of $\sim 8$ from S1 to S2
followed by a dramatic re-brightening by a factor of $\sim 21$ from S2 to S3
(see Table~\ref{tab1}).

\begin{figure}
\includegraphics[angle=-90,scale=0.60]{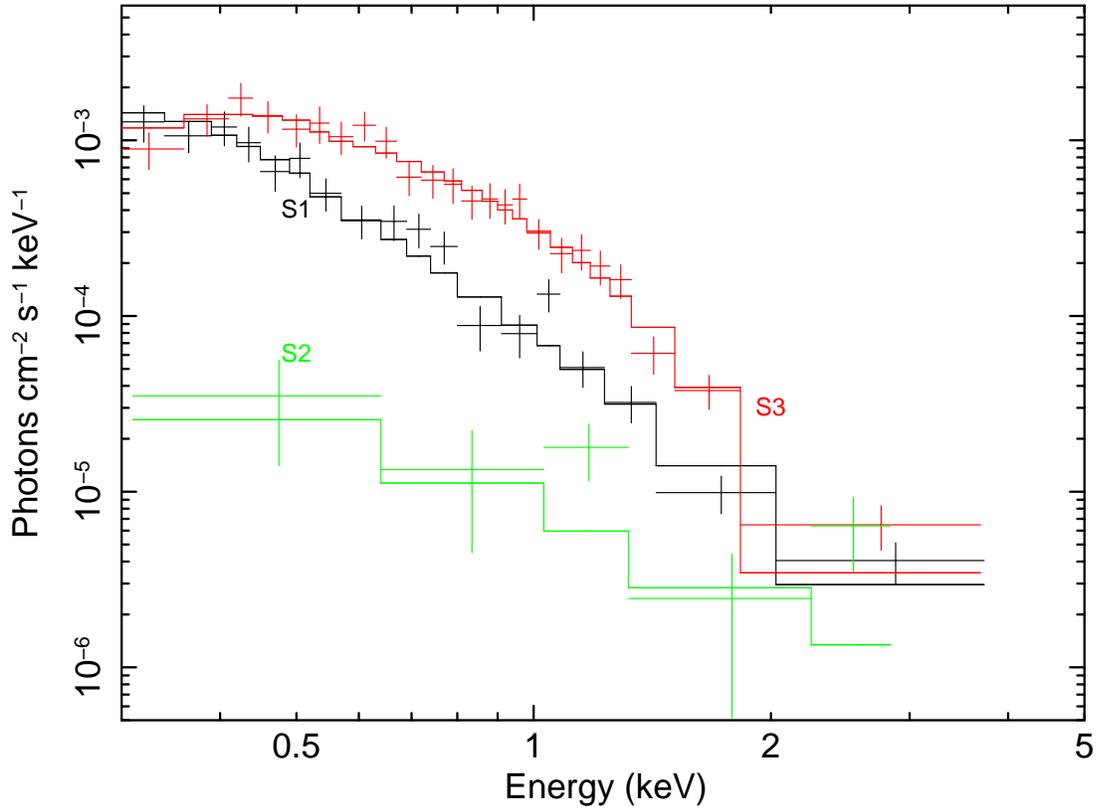}
\caption{{\it Swift}-XRT PC grade 0-12 unfolded spectra of HLX-1: S1 (black)
  and S3 (red).  The solid lines correspond to the best-fit models (see
  Table~\ref{tab1})\label{fig2}. For comparison, we plot the S2 spectrum
  (green) using a power-law with a fixed value of the photon index $\Gamma =
  2$. We used the {\scriptsize XSPEC} command {\scriptsize SETPLOT REBIN 7 7}
  to improve the visual aspect of the S2 spectrum.}
\end{figure}

Fig.~\ref{fig2} shows the unfolded spectra for S1 (black), S2 (green) and S3
(red).  The S1 and S3 spectra were binned to contain a minimum of 20 counts
per channel, and were fitted using the $\chi^2$ minimisation technique. For
all spectra, we fixed the absorption column at $4\times 10^{20}$ cm$^{-2}$,
the best constraint on $N_H$ derived by Farrell et al. (2009).  All the
best-fit models and spectral parameters are given in Table~\ref{tab1}.  From
Fig.~\ref{fig2}, it is immediately obvious that the source has varied
significantly in flux.  The results from S1 when fitted using an absorbed
power-law appear to be consistent with those of XMM1 \citep{Farrell09} which
showed a steep power-law, but are inconsistent with those of XMM2
($\chi^2$/dof=134/18 using the XMM2 best fit model) even though the two
observations were performed only a month apart.  The addition of a disk
black-body component to the power-law model does not significantly improve the
fit ($\Delta \chi^2/dof = 3.1/2$; which corresponds to a F-test probability of
30\%). The use of a unique absorbed disk black-body does not give a good fit
either ($\chi^2/dof = 40.7/16$).  The S3 spectrum is extremely soft. Within
the statistics of the XRT data, the S3 spectrum is better fitted ($\chi^2/dof
= 14.4/21$) using a unique absorbed soft thermal emission than an
absorbed power-law  ($\chi^2/dof = 55.4/21$) with a $N_H$-value fixed at
$4\times 10^{20}$ cm$^{-2}$.  The sparse counts in S2 (28 counts in the
background-subtracted spectrum) prevent us from performing any meaningful
fitting analysis.  Instead, we prefer to use spectral hardness-ratios (see
below). Nevertheless, folding through the {\it Swift}-XRT response kernel the
S1, XMM2 and S3 best-fit models multiplied by a constant factor gives adequate
fits to the S2 data with a constant factor of $\sim 8.2\times 10^{-2}$ ($L\sim
8.1\times 10^{40}$ erg s$^{-1}$) for S1, $\sim 8.3\times 10^{-2}$ ($L\sim
5.0\times 10^{40}$ erg s$^{-1}$) for XMM2 and $\sim 3.1\times 10^{-2}$ ($L\sim
3.4\times 10^{40}$ erg s$^{-1}$) for S3.  The numbers in parentheses are the
corresponding unabsorbed 0.2-10 keV luminosity. In all cases, the S2
luminosity appears to be much lower than those derived for XMM1, XMM2, S1 and
S3 by at least a factor of 10. For comparison, we plot in Fig.~\ref{fig2} the
S2 spectrum (green) using a power-law with a fixed value of the photon index
$\Gamma = 2$.

The fitting results are suggestive of a series of spectral changes.  To
investigate this, we compared the total number of counts in three energy
bands: 0.3-1 keV, 1-3 keV and 3-10 keV (see Table~\ref{tab2}).  There is a
significant drop (by a factor of $\sim 16$) in the soft 0.3-1 keV band between
S1 and S2 when compared to the 1-3 keV band. This suggests a hardening of
the spectrum between S1 and S2.  The source was not detected above 3 keV in
either the S2 or S3 data due to an overall low flux level  in S2 and a
steep spectral slope in S3. From S2 to S3, there is a significant
increase in counts (by a factor of $\sim 44$) in the 0.3-1 keV band when
compared to the other bands, indicating that the X-ray spectrum has softened
again (see also Table~\ref{tab1}).

\section{State transitions in a ULX?}

The detection of the same spectral states as those observed in the GBHBs would
be extremely valuable in constraining the nature of ULXs.  Spectra of other
ULXs have been either interpreted as the source being in a low/hard state or
in a very high state, even if the distinction between the two states is
problematic \citep[e.g.][]{Soria08}. It appears that ULXs in the high/soft or
thermal state are rare. Thermal components with temperatures in the range kT =
1 - 1.5 kev have been observed in the spectra of some ULXs for an emitted
luminosity $L_X < 10^{39}$ erg s$^{-1}$, placing them at the extreme end of
stellar mass BHs \citep{9}.  Winter et al. (2006) claimed the identification
of several ULXs in the thermal state based on the detection of disk black-body
(DBB) spectral components. However, in all cases, the DBB contribution was not
significant with respect to the power-law component. Moreover, no state
transition as seen in GBHBs has been observed in ULXs even those showing large
luminosity variability \citep[e.g.][]{GR09,F08}.  Claims for spectral state
transitions, different from those seen in GBHBs have already been reported
\citep[see e.g.][]{liu02,SM04}. In these papers, the authors claimed evidence
for a high/hard to low/soft transition. Recently, Isobe et al. (2009) claimed
evidence for a spectral transition in the ULX, NGC 2403 Source 3 from a
slim-disk state dominated by a $\sim 1$ keV disk black-body to a very high
state dominated by a power-law spectrum.

\begin{figure}
\includegraphics[angle=0,scale=0.9]{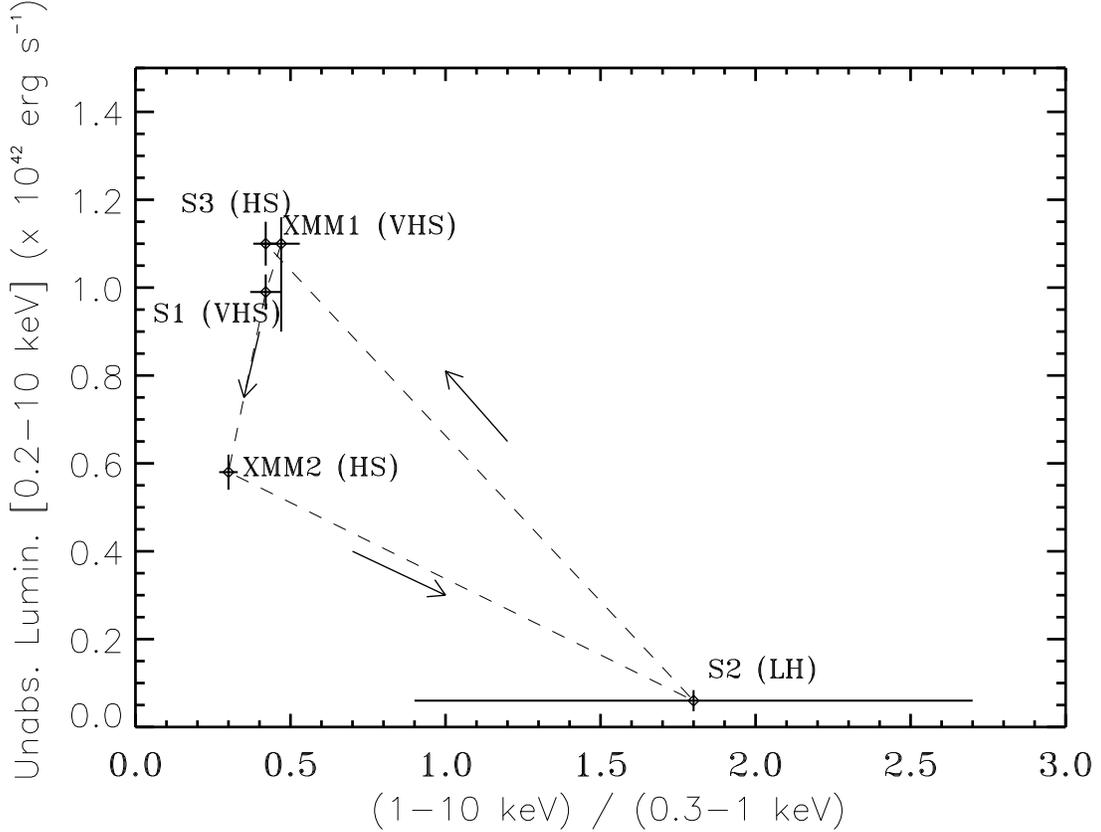}
\caption{Hardness intensity diagram using the {\it Swift} and XMM-Newton
data. The XMM1 and XMM2 points were computed by convolving with the XRT
spectral response the best-fit models in XMM1 and XMM2, respectively. All the
errors are $1\,\sigma$ errors. The S2 0.2-10 keV unabsorbed luminosity was
derived by folding through the {\it Swift}-XRT response kernel the S1, XMM2
and S3 best-fit models multiplied by a constant factor (see Section 2) and
taking the lowest ($L\sim 3.4\times 10^{40}$ erg s$^{-1}$) and highest ($L\sim
8.1\times 10^{40}$ erg s$^{-1}$) values.
\label{fig4}}
\end{figure}

When compared to other ULXs, HLX-1 is truly remarkable not only because
its huge luminosity enables us to claim that it may harbour an IMBH with a
$>500$\,M$_\odot$ mass \citep{Farrell09}, but also because its
luminosity-spectral variability as observed in X-rays (see Section 2) shows
compelling evidence for spectral variation on short timescales (see
Fig.~\ref{fig4}) that are consistent with what has been observed in
GBHBs like GRS 1915+105 \citep{Fender}. Indeed, HLX-1 was likely
to be in the very high state in XMM1 \& S1, while the source was in the
high/soft state in XMM2 with a DBB luminosity corresponding to 80\% of the
total X-ray luminosity. In this case, the transition has occurred in a 1 month
window. In S2, the data suggest that the source may have been in the low/hard
state. The re-brightening and spectral softening in S3 suggest that HLX-1 has
returned to a high/soft state brighter than during XMM2, the transition having
occurred on a relatively short timescale ($< 7$ days).

 The low-temperature of the DBB component measured in XMM2 and S3 is in the
same range as those measured in some ULXs \citep[$\sim 0.1-0.2$ keV;
see][]{Soria08}. This thermal component is often interpreted as direct
emission from the accretion disk. The DBB normalisation $K(\mathrm{XMM2}) \sim
29$ and $K(\mathrm{S3})\sim 14.3$ (via the inner radius of the accretion disk
$R_{in}$) could then be used to constrain the BH mass. To do so, we assumed
that $R_{in}$ corresponds to the radius of the last stable orbit around a
non-rotating BH or a rotating BH with a maximum angular momentum (i.e. $R_{in}
= 0.5-3~R_S \sim 1.5-9\,\mathrm{km}~\left(\frac{M}{M_{\odot}}\right)$ with
$R_S = \frac{2GM}{c^2}$, the Schwarzschild radius where $M$, $c$ and $G$ are
the BH mass, the speed of light and the gravitational constant,
respectively). We found that the derived masses are in the range of IMBH
masses with $M \sim 5.7\times 10^3 - 3.4\times
10^4~M_{\odot}~(\cos({\theta}))^{-\frac{1}{2}}$ from XMM2 and $M \sim 4\times
10^3 - 2.4\times 10^4~M_{\odot}~(\cos({\theta}))^{-\frac{1}{2}}$ from S3,
where $\theta$ is the inclination of the disk with respect to the line of
sight.

Within the limited statistics of the XRT spectrum, we have attempted to fit
the highest quality spectrum (S3) with the slim disk model (Kawaguchi
2003). In order to reduce the number of free parameters, we freezed the source
distance to 95 Mpc and the viscosity parameter to three possible values (0.01,
0.1 and 1). The fits suggest a BH mass larger than $10^3$ M$_{\odot}$, thus
adding another piece of evidence for the presence of an IMBH in HLX-1.

\section{Conclusion}

 We have presented the first evidence for spectral state transitions in HLX-1,
similar to those seen in GBHBs, strengthening the case for a black hole in the
system. Since multi-wavelength observations (e.g.  IR, radio) of HLX-1
continue to exclude alternative explanations for HLX-1, such as a foreground
neutron star, or a background narrow line Seyfert 1 galaxy (Webb et al., in
preparation), HLX-1 still provides the strongest evidence for the existence of
IMBHs in the Universe.

\acknowledgments

 O.G. acknowledges funding from the CNRS and CNES. S.A.F. acknowledges STFC
 funding.  We also thank Cole Miller for useful discussions.

{\it Facilities:} \facility{XMM-Newton}, \facility{Swift}.

\begin{table*}
\begin{center}
\caption{Summary of the X-ray spectral parameters for the best model fits.  \label{tab1}}
\begin{tabular}{cccccc}
\tableline\tableline
Observation & Observation$^a$ & Model & Spectral & $L_X$$^{b}$ & $\chi^2/dof$\\
Number      &  starting date    &  & parameters  & (0.2-10 keV) & \\
            &  &  &  & ($\times 10^{41}$ erg s$^{-1}$) & \\
\tableline
S1 & 2008-10-24 & ABS PL$^c$ & $\Gamma = 3.4\pm 0.2$ &
$9.9\pm 0.5$ & 20/16 \\

  &  &   & $N_H^d = N_H^{\mathrm{XMM2}}$ &  &\\
\tableline
S3 & 2009-08-16 & ABS DBB$^c$  & $kT = 0.26\pm 0.02$ keV & $11.0\pm 1.0$ & 14.4/21 \\
   &            &             & $N_H^d = N_H^{\mathrm{XMM2}}$ & & \\
\tableline
\end{tabular}
\tablenotetext{a}{The S1 observation was performed between 2 {\it XMM-Newton}
  observations: XMM1 $=$ 2004-11-23 and XMM2 $=$ 2008-11-28.} 

\tablenotetext{b}{The unabsorbed 0.2-10 keV luminosity was computed assuming a
source distance of 95.5 Mpc and using the WMAP cosmology.  }

\tablenotetext{c}{ABS PL $=$ absorbed power-law; ABS DBB $=$ absorbed disk black-body.}

\tablenotetext{d}{$N_H^{\mathrm{XMM2}} = 4\times 10^{20}$ cm$^{-2}$ is the
    best constraint on $N_H$ we derived from the XMM2 data
    \citep{Farrell09}. It is the sum of the Galactic column absorption in
    the direction of HLX-1 $N_H^{Gal} = 2\times 10^{20}$ cm$^{-2}$
    \citep{DL90} and the intrinsic column absorption along the line of
    sight.}

\end{center}
\end{table*}

\begin{table*}
\begin{center}
\caption{Comparison of the number of counts in 3 energy bands: 0.3-1 keV, 1-3
  keV and 3-10 keV. The last column gives the source count rate in the 0.3-10
  keV band. All the numbers were derived using a 20 pixel radius circle for
  the source and the background. \label{tab2}}
\begin{tabular}{cccccc}
\tableline\tableline
Observation & Exposure & Counts $^a$ & Counts$^a$ & Counts$^a$ & Count rate\\
Number      &  time    & (0.3-1 keV)& (1-3 keV)  & (3-10 keV)  & (0.3-10 keV)\\
            & (ks) &  &  & & (cts s$^{-1}$)\\
\tableline
S1 & 33.45 & $274 \pm 17$ & $97 \pm 10$ & $18\pm 5$ & $1.2\times 10^{-2}$ \\
  & (19.17) & ($157 \pm 13$)$^b$ & ($56\pm 7$)$^b$ & ($10\pm 4$)$^b$ & \\
\tableline
S2 & 19.17 & $10\pm 4$ & $18\pm 5$ & $0^{+2}_{-0}$ & $1.5\times 10^{-3}$\\
\tableline
S3 & 19.09 & $437\pm 21$ & $185\pm 14$ & $0^{+2}_{-0}$ & $3.3\times 10^{-2}$ \\
   & (19.17) & ($438 \pm 21$)$^b$ & ($186 \pm 14$)$^b$ & --&\\
\tableline

\end{tabular}
\tablenotetext{a}{The errors quoted above are 1 $\sigma$ error. When the
  number of counts in a given energy band is less than 20, the 1-$\sigma$
  errors were computed using the following formula: $\sigma = 1+\sqrt{N+0.75}$
  instead of $\sigma = \sqrt{N}$ \cite{G86}.  } 

\tablenotetext{b}{The number of counts are computed using an exposure time of
  19.17 ks to facilitate the comparison between all the {\it Swift}
  observations.  }

\end{center}
\end{table*}

\end{document}